\documentclass[aps,graphicx,12pt]{revtex4}
\usepackage{amssymb}
\usepackage{amsmath}
\usepackage{amscd}
\usepackage{subfigure}
\usepackage{graphicx}

\begin{document}

\title[Short Title]{Driving three atoms into singlet state in an optical cavity
via adiabatic passage of dark state}

\author{Mei Lu$^{1}$}
\author{Yan Xia$^{1,2,}$
\footnote{E-mail: xia-208@163.com}}
\author{Jie Song$^{3}$}
\author{He-Shan Song$^{2}$}

\affiliation{$^{1}$Department of Physics, Fuzhou University, Fuzhou
50002, China\\$^{2}$School of Physics and Optoelectronic Technology,
Dalian university of Technology, Dalian 116024,
China\\$^{3}$Department of Physics, Harbin Institute of Technology,
Harbin 150001, China}

\begin{abstract}In this paper, we propose an efficient scheme
to drive three atoms in an optical cavity into singlet state via
adiabatic passage. Appropriate Rabi frequencies of the classical
fields are selected to realize present scheme. The scheme is robust
against the deviations in the pulse delay and laser intensity
through some simple analysis of adiabatic condition. It is notable
that the estimated range of effective adiabaticity condition
coincides with the numerical results. When taking dissipation into
account, we show that the process is immune to atomic spontaneous
emission as the atomic excited states are never populated in
adiabatic evolution. Moreover, under certain conditions, the cavity
decay also can be efficiently suppressed.
\end{abstract}

\maketitle

\section{INTRODUCTION}
The technique of adiabatic passage
\cite{jr-pra741,ug-jcp363,kb-rmp003} has been proved as an effective
coherently control dynamical process to realize quantum information
processing (QIP) in many schemes
\cite{zk-pra318,hg-pra305,sbz-prl502,xl-pra321,zjd-pra303,js-jpb503,
zby-cpb205,js-apl102}. Unlike the stimulated Raman adiabatic passage
(STIRAP) where the Stokes pulse vanishes first, the two pulses
vanish simultaneously while maintaining a constant finite ratio of
amplitudes in f-STIRAP \cite{nv-jpb535,ma-pra805,ma-pra339}, which
guarantees the creation of any pre-selected coherent superposition
of ground states. A remarkable superiority of STIRAP and f-STIRAP is
that if the evolution is adiabatic, the states of the system evolve
within an adiabatic dark state subspace and the relevant states
contain no contribution of the excited atomic states, so the
spontaneous emission from excited states can be suppressed. Another
advantage of adiabatic passage is the simpleness, for it needs not
consider the precise tuning of pulse areas, pulse widths, pulse
shapes, pulse delay and detunings.

Cabello has proposed a special type of entangled state which is
called $N$-particle $N$-level singlet states in 2002
\cite{ca-prl402}. These states are key solutions to many problems
such as ¡®$N$-strangers¡¯, ¡®secret sharing¡¯ and ¡®liar
detection¡¯, which have no classical solutions. It has been shown
that some types of supersinglets may violate the local hidden theory
\cite{mnd-prd356} and can be used to construct decoherence-free
subspaces (DFSs), which are robust against collective decoherence
\cite{ca-jmo049}. Nevertheless, the generation of these states for
$N=3$ which have the form
\begin{eqnarray}\label{e1}
\frac{1}{\sqrt{6}}(|012\rangle-|102\rangle-|210\rangle+|120\rangle
+|201\rangle-|021\rangle),
\end{eqnarray}
is still a problem both in theory and experiment. Here $|0\rangle,\
|1\rangle$ and $|2\rangle$ represent three ground states. Jin
\emph{et al.} have proposed a scheme of generating a supersinglet of
three three-level atoms in microwave cavity QED based on the
resonant atom-cavity interaction \cite{jgs-pra307}. The scheme is
sensitive to the loss of cavity since the three atoms are
sequentially sent through three different cavities and cavity fields
which act as memories. Lin \emph{et al.} also have raised a protocol
for the preparation of a singlet state with three atoms via Raman
transitions \cite{lgw-pra308}. However, there will be a considerable
influence caused by the cavity decay and spontaneous emission of the
atoms. Shao \emph{et al.} have put forward an approach by converting
two-atom singlet state into three-atom singlet state via quantum
Zeno dynamics \cite{xqs-njop040}, which needs to control the
interaction time exactly.

To overcome these problems in Refs.
\cite{jgs-pra307,lgw-pra308,xqs-njop040}, we propose a scheme where
the state of the system evolves within a dark-state subspace via
adiabatic passage. Three Gaussian shape pulses are used to complete
the scheme and by this way the two Rabi frequencies of the laser
fields maintain a constant finite ratio of amplitudes to get the
desired final singlet states. Also sufficient adiabaticity can be
achieved by choosing such pulses, for the estimates based on simple
analysis of adiabaticity condition show the scheme is robust against
deviations such as the pulse delay and pulse intensity, and the
analysis is also validated by numerical calculation. Moreover, we
also analyze the influence of the choice of laser intensity on the
dissipation and achieve a relative high fidelity by choosing
appropriate parameters.

The paper is organized as follows. In section II, we introduce the
models for adiabatic passage and the pulses to realize the scheme.
In section III, we discuss the robustness of parameters mismatch and
dissipation due to the atomic spontaneous emission and cavity delay.
Section IV contains the concluding remarks.

\section{Generation of three-atom singlet state}
As depicted in Fig. 1, three four-level atoms with tripod
configuration are trapped in a bimodal vacuum cavity field. Each
atom has an excited state $|e\rangle$ and three ground states
$|f_L\rangle$, $|f_R\rangle$ and $|r\rangle$. Supposed that the
transition between the levels
$|e\rangle_i\leftrightarrow|f_L\rangle_i(|f_R\rangle_i)\ (i=1,2,3)$
is resonantly coupled to the cavity mode with the coupling strength
$g_{iL}(g_{iR})$ and the transition
$|e\rangle_i\leftrightarrow|r\rangle_i$ is resonantly driven by the
classical pulse with the Rabi frequency $\Omega_i$. In the
interaction picture, the Hamiltonian for the whole system can be
written as $(\hbar=1)$
\begin{eqnarray}\label{e2}
H_{tot}&=&H_l+H_c,\cr\cr
H_l&=&\sum^3_{i=1}\Omega_i(|e\rangle_i\langle r|+|r\rangle_i\langle
e|),\cr\cr H_c&=&\sum^3_{i=3}g_{iL}(a_L|e\rangle_i\langle
f_L|+a^\dag_L|f_L\rangle_i\langle e|) + g_{iR}(a_R|e\rangle_i\langle
f_R|+a^\dag_R|f_R\rangle_i\langle e|),
\end{eqnarray}
where $a^\dag_L(a^\dag_R)$ and $a_L(a_R)$ are the creation and
annihilation operations for the left(right)-circular polarization
cavity mode, respectively. We assuming $g_{iL}=g_{iR}=g,\
\Omega_2(t)=\Omega_3(t)=\Omega(t)$, and $\Omega_1(t)$ to be real in
the present paper for simplicity, the whole system will evolve in
the following closed subspace:
\begin{eqnarray}\label{e3}
&&|\phi_1\rangle=|rf_Lf_R\rangle_{123}|00\rangle_{a_La_R},\;
|\phi_2\rangle=|rf_Rf_L\rangle_{123}|00\rangle_{a_La_R},\cr\cr&&
|\phi_3\rangle=|f_Lrf_R\rangle_{123}|00\rangle_{a_La_R},\;
|\phi_4\rangle=|f_Rrf_L\rangle_{123}|00\rangle_{a_La_R},\cr\cr&&
|\phi_5\rangle=|f_Lf_Rr\rangle_{123}|00\rangle_{a_La_R},\;
|\phi_6\rangle=|f_Rf_Lr\rangle_{123}|00\rangle_{a_La_R},\cr\cr&&
|\phi_7\rangle=|ef_Lf_R\rangle_{123}|00\rangle_{a_La_R},\;
|\phi_8\rangle=|ef_Rf_L\rangle_{123}|00\rangle_{a_La_R},\cr\cr&&
|\phi_9\rangle=|f_Lef_R\rangle_{123}|00\rangle_{a_La_R},\;
|\phi_{10}\rangle=|f_Ref_L\rangle_{123}|00\rangle_{a_La_R},\cr\cr&&
|\phi_{11}\rangle=|f_Lf_Re\rangle_{123}|00\rangle_{a_La_R},\;
|\phi_{12}\rangle=|f_Rf_Le\rangle_{123}|00\rangle_{a_La_R},\cr\cr&&
|\phi_{13}\rangle=|f_Lf_Lf_R\rangle_{123}|10\rangle_{a_La_R},\;
|\phi_{14}\rangle=|f_Lf_Rf_L\rangle_{123}|10\rangle_{a_La_R},\cr\cr&&
|\phi_{15}\rangle=|f_Rf_Lf_L\rangle_{123}|10\rangle_{a_La_R},\;
|\phi_{16}\rangle=|f_Rf_Rf_L\rangle_{123}|01\rangle_{a_La_R},\cr\cr&&
|\phi_{17}\rangle=|f_Rf_Lf_R\rangle_{123}|01\rangle_{a_La_R},\;
|\phi_{18}\rangle=|f_Lf_Rf_R\rangle_{123}|01\rangle_{a_La_R},
\end{eqnarray}
the subscripts $1,2,3,a_L$ and $a_R$ represent atom 1, atom 2, atom
3, left-circular cavity mode and right-circular cavity mode,
respectively.

There are six dark states with null eigenvalue in this subspace. We
orthogonalize these states and get a special dark state $|S\rangle$
which will evolve into an independent subspace while other states
remain unchanged. This state can be expressed as
\begin{eqnarray}\label{e4}
|S\rangle=&&\frac{1}{N}\Big[-\Omega(t)
g(|\phi_1\rangle-|\phi_2\rangle)
-\frac{\Omega_1(t)g}{2}(|\phi_3\rangle-|\phi_4\rangle-|\phi_5\rangle+|\phi_6\rangle)
\cr\cr&&+\frac{\Omega_1(t)\Omega(t)}{2}(|\phi_{13}\rangle-|\phi_{14}\rangle
-|\phi_{16}\rangle+|\phi_{17}\rangle)\Big],
\end{eqnarray}
where $N=\sqrt{\Omega_1^2(t)g^2+2\Omega^2(t)g^2
+\Omega_1^2(t)\Omega^2(t)}$. Note that the state in Eq. (4) contains
no contribution from the atomic excited states
\cite{pa-prl095,py-prl788}, which can be considered as unpopulated
during the whole interaction process, if the evolution is adiabatic.
When the system is initially in the state
\begin{eqnarray}\label{e5}
|\phi_i\rangle=\frac{1}{\sqrt{2}}(|\phi_1\rangle-|\phi_2\rangle),
\end{eqnarray}
under the condition
\begin{eqnarray}\label{e6}
g\gg\Omega_1(t),\Omega(t),
\end{eqnarray}
and the Rabi frequencies following such behaviour
\begin{eqnarray}\label{e7}
\lim_{t\rightarrow-\infty}\frac{\Omega_1(t)}{\Omega(t)}=0,\
\lim_{t\rightarrow+\infty}\frac{\Omega_1(t)}{\Omega(t)}=\tan\alpha,
\end{eqnarray}
we have an approximate evolution process from the initially
entangled state $|\phi_i\rangle$ to three-atom singlet state
\begin{eqnarray}\label{e8}
|\phi_t\rangle=\frac{1}{\sqrt{6}}(|\phi_1\rangle-|\phi_2\rangle+|\phi_3\rangle
-|\phi_4\rangle-|\phi_5\rangle+|\phi_6\rangle),
\end{eqnarray}
when $\alpha=\arctan2$, which is the result we need. Let us define
that $\theta(t)=\arctan[\Omega_1(t)/\Omega(t)]$, note that the rate
of the change of the mixing angle $\theta(t)$ must be much smaller
compared to the smallest separation $\Delta\omega(t)$ of the
corresponding eigenvalues \cite{jr-pra741,nv-jpb535}, the specific
expression is
\begin{eqnarray}\label{e9}
|\dot{\theta}(t)|\ll\Delta\omega(t).
\end{eqnarray}
Under these conditions, the evolution is adiabatic, and the system
will remain in the dark state $|S\rangle$.

Next we discuss the Rabi frequencies and other correlative
parameters that make the scheme experimentally feasible. We use
three time-dependent pulses as follows£º
\begin{eqnarray}\label{e10}
&&\Omega_1(t)=\sin\alpha\Omega_0\exp^{-(t-\tau)^2/T^2},
\cr\cr&&\Omega(t)=\cos\alpha\Omega_0\exp^{-(t-\tau)^2/T^2}+\;
\Omega_0\exp^{-(t+\tau)^2/T^2},
\end{eqnarray}
and the variation of the two time-dependent Rabi frequencies
$\Omega_1$ (red dashed curve) and $\Omega$ (blue solid curve) of
lasers for atoms is shown in Fig. 2(a). By choosing $\tau=60/g,\
T=80/g,\ \Omega_0=0.2g$, we can see that when $t\geq100/g$, the two
Rabi frequencies are approximate to meet the relation
$\Omega_1\simeq\frac{1}{2}\Omega$. Thus the initial state
$|\phi_i\rangle$ will transfer into the target state
$|\phi_t\rangle$. Fig. 2(b) shows the time evolution of the
populations of the components of the state $|S\rangle$ $P_1(P_2)$
(green dashed curve), $P_3(P_4,P_5,P_6)$ (purple dotted curve) and
$P_{13}(P_{14},P_{16},P_{17})$ (red solid curve), respectively.

It is obvious that the population is almost completely transferred
from the state $|\phi_i\rangle$ to the state $|\phi_t\rangle$
without populating other states during the dynamical process, which
means the influence of the atomic spontaneous emission is
effectively suppressed. For the density matrix of the two entangle
states $\rho_1,\;\rho_2$, the Bures fidelity can be defined as
\cite{rj-jmo315}
\begin{eqnarray}\label{e11}
F=\Big(tr\sqrt{\rho_1^{1/2}\rho_2\rho_1^{1/2}}\Big)^2,
\end{eqnarray}
The relation between the fidelity and the evolution time $t$ is
shown in Fig. 2(c). When $T=100/g$, the fidelity of the three-single
state is 0.9996, which means we finally realize an almost perfect
target state. Note that we need not control the time accurately
since a fidelity $F$ higher than 0.8775 can be obtain when
$t\geq20/g$.

\section{ADIABATICITY CONDITION AND NUMERICAL ANALYSIS}
Now we come to discuss the conditions that a system with the initial
state $|\phi_i\rangle$ evolves adiabatically into the target state
$|\phi_t\rangle$. In our analysis of the robustness of adiabatic
passage against variations in the experimental parameters, we start
with the adiabatic condition (9). For the restrictive conditions of
Eq. (6) and the given pulse shapes we have
\begin{eqnarray}\label{e12}
\dot{\theta}(t)&=&\frac{4\tau}{T^2}\frac{\xi(t)\sin\alpha}{\sin^2\alpha+[\cos\alpha+\xi(t)]^2},
\cr\cr\Delta\omega(t)&=&\sqrt{\frac{3g^2}{2}-\frac{\sqrt{9g^4+2g^2[\Omega_1^2(t)-\Omega^2(t)]
+[\Omega_1^2(t)-\Omega^2(t)]^2}}{2}+\frac{\Omega_1^2(t)}{2}+\frac{\Omega^2(t)}{2}}
\cr\cr&\simeq&\sqrt{\frac{\Omega_1^2(t)}{3}+\frac{2\Omega^2(t)}{3}}
\cr\cr&=&\Omega_0e^{-(t-\tau)^2/T^2}\sqrt{\frac{\sin^2\alpha}{3}
+\frac{2}{3}[\cos\alpha+\xi(t)]^2}=\Omega_{eff}(t),
\end{eqnarray}
where $\xi(t)=e^{-4\tau t/T^2}$. Then a limit for the pulse delay
can be get by analysing the upper expressions.

We know that non-adiabatic conditions are most likely to occur when
$\dot{\theta}(t)$ reaches a maximum, that is $\xi(t_0)=1$. Take
$t_0=0$ for example, then Eq. (12) is equal to
\begin{eqnarray}\label{e13}
\dot{\theta}(t_0)&=&\dot{\theta}_{max}(t)=\frac{2\tau}{T^2}\tan\frac{1}{2}\alpha,
\cr\cr\Omega_{eff}(t_0)&=&\Omega_0e^{-\tau^2/T^2}\sqrt{\frac{\sin^2\alpha}{3}
+\frac{2}{3}(\cos\alpha+1)^2}.
\end{eqnarray}
The maximum of $\dot{\theta}(t)$ increases with $\tau$ (see Eq.
(13)), while $\Omega_{eff}(t_0)$ is not necessarily the maximum of
$\Omega_{eff}(t)$. To reach the adiabatic conditions (9) we must
have $\Omega_{eff}(t_0)\geq n\dot{\theta}(t_0)$, where $n$ is a
`sufficiently large' number and the choice of which depends on how
much non-adiabaticity can be allowed. Then we find an upper bound on
$\tau$,
\begin{eqnarray}\label{e14}
\Omega_0T\geq\frac{2n\tau}{T}\frac{\tan\frac{1}{2}\alpha}
{\sqrt{\frac{\sin^2\alpha}{3}+\frac{2}{3}(\cos\alpha+1)^2}}e^{\tau^2/T^2}.
\end{eqnarray}
We can see that the Rabi frequency needs to increase exponentially
with $\tau$ to suppress the non-adiabatic transitions.

Note that both $\dot{\theta}(t)$ and $\Omega(t)$ are pulse-shaped,
it is convenient to find their full widths at half maximum(FWHM) for
the following discussions.
\begin{eqnarray}\label{e15}
T_{\dot{\theta}}&\simeq&\frac{T^2}{\tau}\ln\left(\sqrt{1+\cos^2\frac{1}{2}\alpha}
+\cos\frac{1}{2}\alpha\right), \cr\cr
T_{\Omega}&\simeq&2\tau+2T\sqrt{\ln2}.
\end{eqnarray}
When $\tau\rightarrow0$, the pulses of $\dot{\theta}(t)$ will be
broaden and get broader than $\Omega(t)$ . Then condition (9) will
be violated in the early time and late time during the evolution.
This problem can be solved by controlling the width of
$\dot{\theta}(t)$ smaller than the width of $\Omega(t)$, that is
$T_{\dot{\theta}}\leq T_{\Omega}$. Then we obtain a lower bound for
$\tau$. When $\alpha=\arctan2$ it reads as
\begin{eqnarray}\label{e16}
\tau\geq0.25T.
\end{eqnarray}

From the above analyses we can see that although for any
sufficiently strong laser pulse and delay $\tau>0$ the f-STIRAP
should work, there exists a more accurate adiabatic condition. For
$n=5$, the range is $0.25T\leq\tau\leq0.97T$ for $\Omega_0T=12$,
$0.25T\leq\tau\leq1.06T$ for $\Omega_0T=16$ and
$0.25T\leq\tau\leq1.13T$ for $\Omega_0T=20$. We also plot the
relationship of the fidelity $F$ versus the ratio $\tau/T$ in the
situation that $\Omega_0T=12$ (blue dotted curve), $\Omega_0T=16$
(green solid curve) and $\Omega_0T=20$ (red dashed curve) by solving
the master equation numerically in Fig. 3(a). The figure coincides
with the results we deduce from Eq. (9) almost perfectly, which
verifies that the adiabaticity is most easily achieved under the
range we obtain. Fig. 3(a) also implies that for a particular value
of $\Omega_0T$, there exists a relative wide range of $\tau/T$,
which means that our protocol is robust against the pulse delay in
practical experiment. Fig. 3(b) shows the change of the time
dependence of the smallest separation $\Delta\omega(t)$ (blue dotted
curve), the simplified smallest separation $\Omega_{eff}(t)$ (green
solid curve) and the mixing angle $\theta(t)$ (red dashed curve)
with the chosen parameters, which implies that the parameters our
scheme chooses fit condition (9) well.

In all the above discussions, we have not considered any dissipation
and assume the system does not interact with the environment.
However, the system will interact with the environment inevitably
which has influence on the availability of our scheme. Hence we will
focus on discussing the influence of dissipation induced by the
atomic spontaneous emission and the cavity decay. When we consider
decoherence, the master equation of motion for the density matrix of
the whole system can be expressed as
\begin{eqnarray}\label{e17}
\dot{\rho}&=&-i[H_{tot},\rho]-\frac{\kappa_L}{2}(a_L^\dag
a_L\rho-2a_L\rho a_L^\dag+\rho a_L^\dag
a_L)\cr\cr&-&\frac{\kappa_R}{2}(a_R^\dag a_R\rho-2a_R\rho
a_R^\dag+\rho a_R^\dag
a_R)\cr\cr&-&\sum^3_{k={1}}\sum_{m=f_L,g,f_R}\frac{\Gamma_{em}^k}{2}
(\sigma_{em}^k\sigma_{me}^k\rho-2\sigma_{me}^k\rho\sigma_{em}^k
+\rho\sigma_{em}^k\sigma_{me}^k),
\end{eqnarray}
where $\Gamma_{em}^k$ is the spontaneous emission rate from the
excited state $|e\rangle$ to the ground states $|m\rangle$
($m=f_L,r,f_R$) of the $k$th atom. $\kappa_L(\kappa_R)$ is the decay
rate of the left(right)-circular cavity mode. We assume
$\Gamma_{em}^k=\Gamma=\Gamma_0/3$ and $\kappa_L=\kappa_R=\kappa$ for
simplicity. Fig. 4 shows the relationships of the fidelity $F$
versus the ratios $\Omega_0/g$ and $\Gamma/g$, $\Omega_0/g$ and
$\kappa/g$, respectively. The spontaneous emission rate makes a
slighter influence on $F$ under a larger laser intensity while
cavity decay causes an opposite situation. The physical mechanism
behind the behave of the former case is just the adiabatic condition
(9), which is satisfied better along with laser intensity
increasing. That means under a relative large laser intensity, i.e.
at $\Omega_0/g=0.3$ in Fig.4., the passage is more likely to evolve
within the adiabatic dark state subspace which does not involve the
excited atomic state $|e\rangle$. Therefore, the dissipation caused
by atomic spontaneous emission decreases. However the populations of
states where the cavity field are excited increase with laser
intensity according to Eq. (4), which increase the dissipation
caused by cavity decay finally. In such a way, the change of laser
intensity decreases one error source while increasing another. So an
appropriate value $\Omega_0$ should be chosen when taking both the
two factors into account. We also plot the relationship of the
fidelity $F$ versus the ratios $\kappa/g$ and $\Gamma/g$ by solving
the master equation numerically in Fig. 5. Therefore we can see that
under certain conditions both atomic spontaneous emission and cavity
decay have a slight influence in fidelity $F$, since for a large
atomic spontaneous emission $\Gamma/g=0.05$ and cavity decay
$\kappa/g=0.05$, the fidelity is still about 0.9244. Therefore our
scheme is robust against the two error sources and achieve a
superior result in theory.

Finally, we give a brief discussion about the basic factors for the
experimental realization. The atomic configuration might be achieved
in cesium atoms in our scheme. The state $|r\rangle$ corresponds to
$F=4, m=3$ hyperfine state of $6^2S_{1/2}$ electronic ground state,
$|f_L\rangle$ corresponds to $F=3, m=2$ hyperfine state of
$6^2S_{1/2}$ electronic ground state, $|f_R\rangle$ correspond to
$F=3, m=4$ hyperfine state of $6^2S_{1/2}$ electronic ground state,
$|e\rangle$ corresponds to $F=4, m=3$ hyperfine state of
$6^2P_{1/2}$ electronic state, respectively. In recent experimental
condition \cite{sm-pra817,jr-pra806}, the parameters
$g=2\pi\times750MHz,\ \Gamma_0=2\pi\times2.62MHz,\
\kappa=2\pi\times3.5MHz$ and the optical cavity mode wavelength in
the range between $630\sim850$ nm is predicted to achieve. By
substituting the ratios $\kappa/g=0.0047,\Gamma/g=0.0035$ into Eq.
(17), we will obtain a high fidelity of about 0.956, which shows our
scheme to prepare three-atom singlet state $|\phi_t\rangle$ is
relatively robust against a realized one.

\section{CONCLUSION}
In summary, we have proposed a scheme to generate a three-single
state for three atoms trapped in an optical cavity via the adiabatic
passage of dark state. The significant feature is that we need not
to control the laser time exactly and it is robust against
variations in the laser parameters such as pulse delay and laser
intensity. So the scheme is robust, effective and simple. When
considering dissipation, we can find that the protocol is robust
against atomic spontaneous emission since the states evolve in a
closed subspace where the atoms remain in the ground states in a
adiabatic evolution. By choosing proper parameters, the scheme is
also insensitive to cavity decay  by numerical calculation
intuitionally. The result shows that the scheme have a high fidelity
and may be possible to implemented with the current experiment
technology.

\acknowledgements

This work is supported by the funds from Education Department of
Fujian Province of China under Grant No. JB08010, No. JA10009 and
No. JA10039, the National Natural Science Foundation of Fujian
Province of China under Grant No. 2009J06002 and No. 2010J01006, the
National Natural Science Foundation of China under Grant No.
11047122 and No. 11105030, Doctoral Foundation of the Ministry of
Education of China under Grant No. 20093514110009, and China
Postdoctoral Science Foundation under Grant No. 20100471450.

\newpage
\begin{figure}
\centering\scalebox{0.8}{\includegraphics [width=10cm]{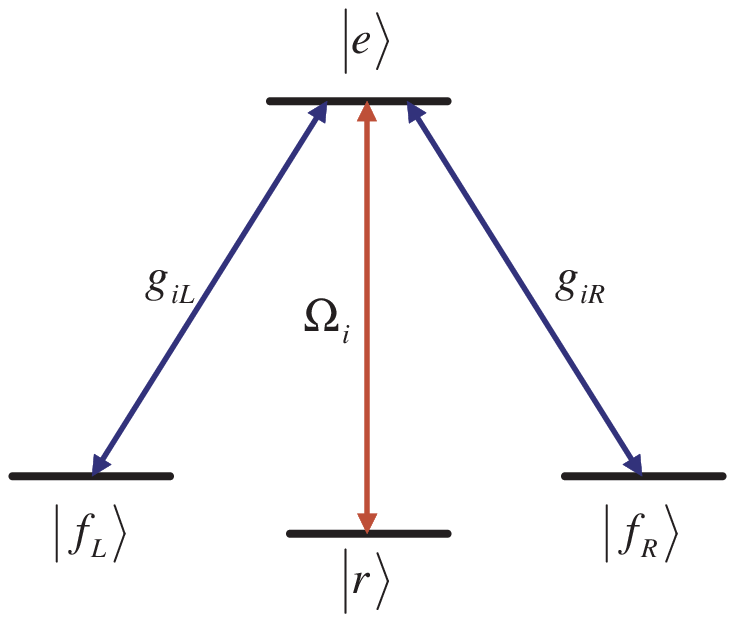}}
\caption{The level configuration of the scheme. The transition
$|e\rangle\rightarrow|f_L\rangle$ and
$|e\rangle\rightarrow|f_R\rangle$ are coupled to left-circularly and
right-circularly polarised cavity modes, respectively. A classical
laser driver the transitions $|r\rangle\rightarrow|e\rangle$.}
\label {Fig.1}
\end{figure}

\begin{figure}
\centering\subfigure[]{\label{Fig.sub.a}\includegraphics[width=10cm]{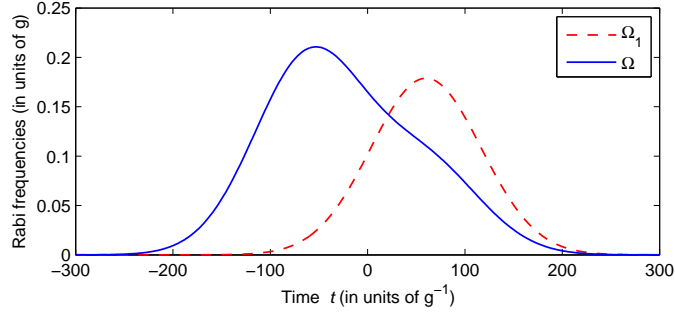}}
\subfigure[]{\label{Fig.sub.b}\includegraphics[width=10cm]{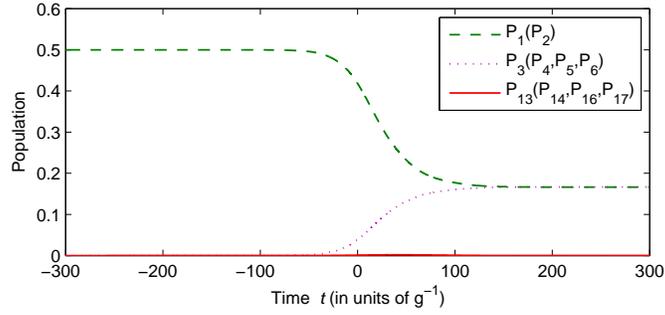}}
\subfigure[]{\label{Fig.sub.c}\includegraphics[width=10cm]{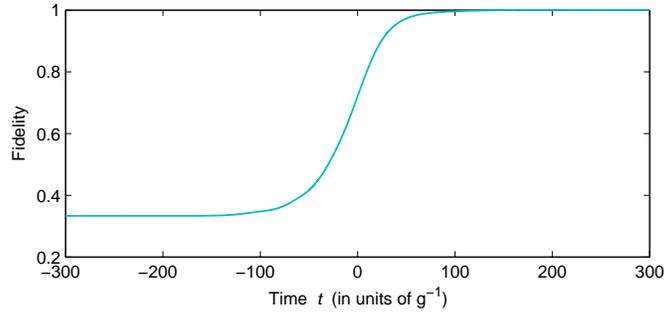}}
\caption{(a) The time dependence of the laser fields for atoms. Here
the red dashed curve and the blue solid curve represent the Rabi
frequencies $\Omega_1$ and $\Omega$, respectively. (b) Time
evolution of the populations. Here the green dashed curve, the
purple dotted curve and the red solid curve represent the
populations $P_1(P_2)$, $P_3(P_4,P_5,P_6)$ and
$P_{13}(P_{14},P_{16},P_{17})$, respectively. (c) Time evolution of
the fidelity. We have chosen $\tau=60/g,\ T=80/g,\ W_2/g=100,\
\Omega_0=0.2g$.}\label{Fig.2}

\end{figure}

\begin{figure}
\centering\subfigure[]{\label{Fig.sub.a}\includegraphics[width=10cm]{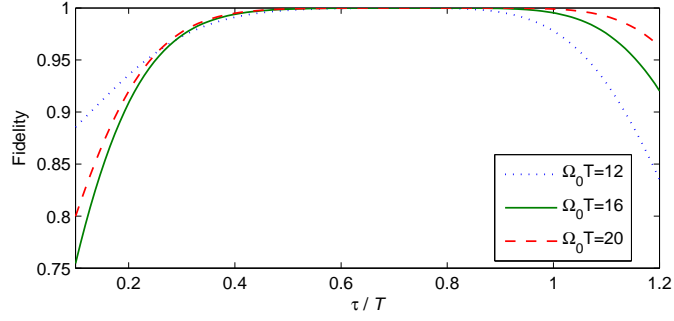}}
\subfigure[]{\label{Fig.sub.b}\includegraphics[width=10cm]{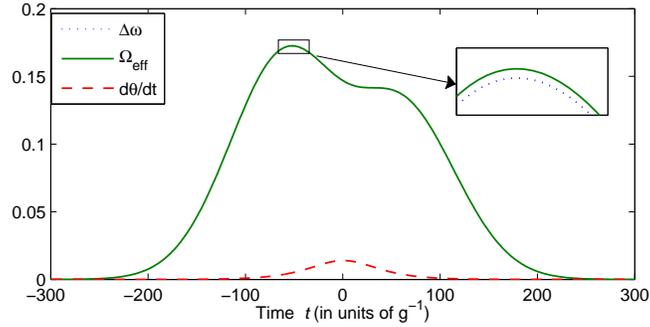}}
\caption{(a) The fidelity $F$ vs. the ratio $\tau/T$ when
$\Omega_0T$ is in different values. Here the blue dotted curve, the
green solid curve and the red dashed curve represent the values
$\Omega_0T=12,16$ and 20, respectively. (b) The time dependence of
the the change of the smallest separation $\Delta\omega$, the
simplified smallest separation $\Omega_{eff}$ and the mixing angle
$\theta$. Here the blue dotted curve, the green solid curve and the
red dashed curve represent the change of $\Delta\omega$,
$\Omega_{eff}$ and $\theta$, respectively. We have chosen
$\tau/g=60,\ W/g=80,\ \Omega_0/g=0.2$.}\label{Fig.3}
\end{figure}

\begin{figure}
\centering\subfigure[]{\label{Fig.sub.a}\includegraphics[width=8cm]{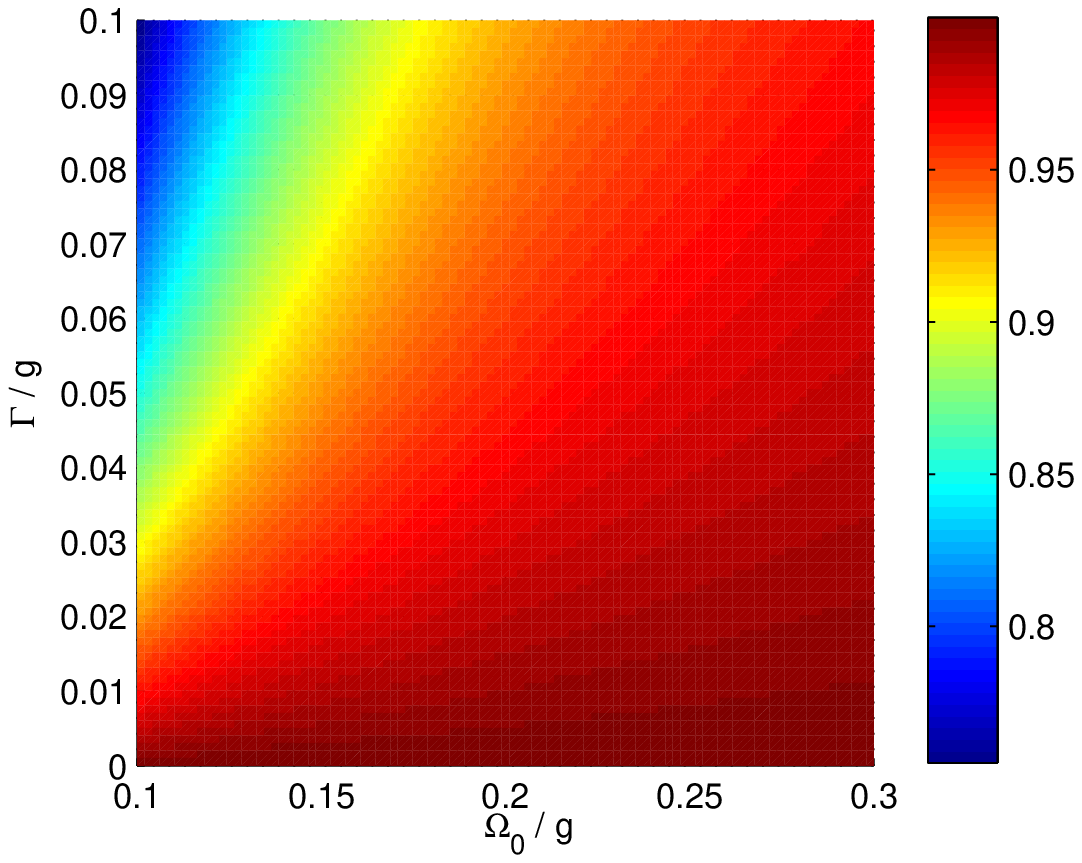}}
\subfigure[]{\label{Fig.sub.b}\includegraphics[width=8cm]{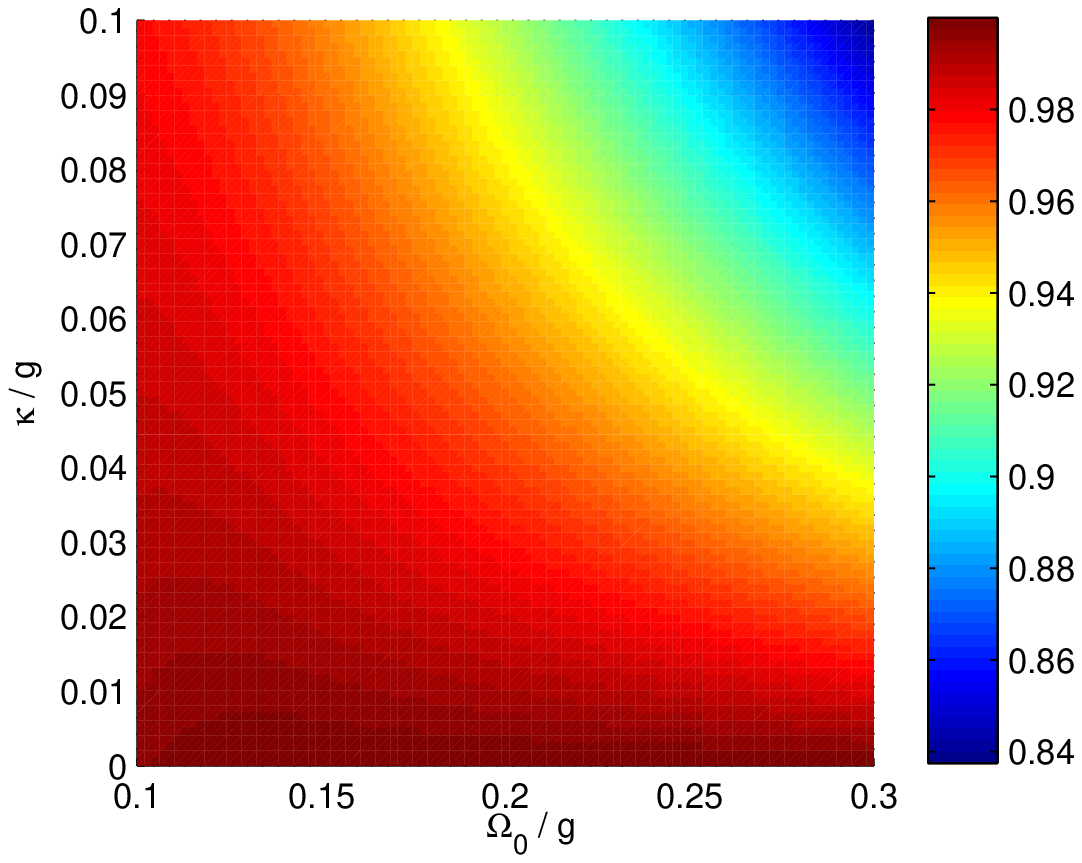}}
\caption{(a) The fidelity $F$ vs. the ratios $\Omega_0/g$ and
$\Gamma/g$. (b) The fidelity $F$ vs. the ratios $\Omega_0/g$ and
$\kappa/g$.}\label{Fig.4}
\end{figure}

\begin{figure}
\scalebox{0.8}{\includegraphics {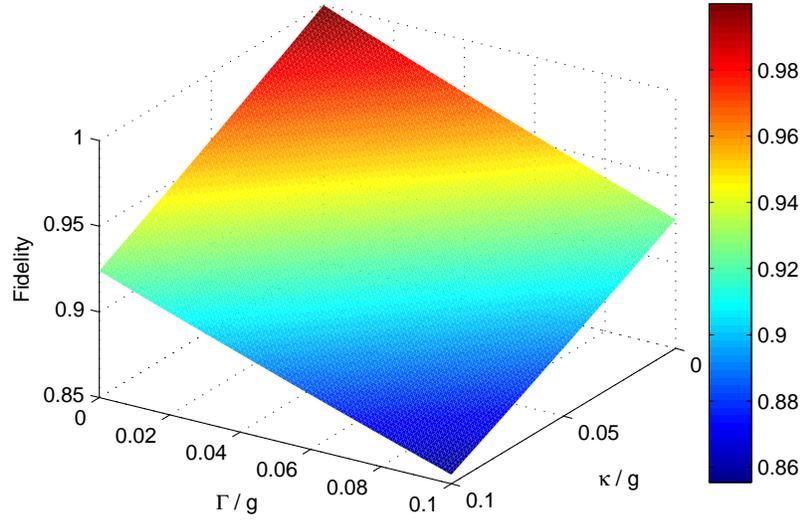}} \caption{The infuence of
ratios $\kappa/g$ and $\Gamma/g$ on the Fidelity $F$ of the
three-atom singlet state.} \label {FIG.5}
\end{figure}

\end{document}